\documentclass[review]{elsarticle}
\usepackage{lineno,hyperref}
\usepackage{comment}
\modulolinenumbers[5]
\journal{Journal of \LaTeX\ Templates}

\begin{document}

\begin{frontmatter}

\title{Search for $\eta$ mesic $^3$He with the WASA-at-COSY facility in the \mbox{$pd\rightarrow$ $^{3}\hspace{-0.03cm}\mbox{He} 2\gamma$} and  $pd\rightarrow$ $^{3}\hspace{-0.03cm}\mbox{He} 6\gamma$ reactions}

\author[IKPUU]{P.~Adlarson}
\author[ASWarsN]{W.~Augustyniak} 
\author[IPJ]{W.~Bardan}     
\author[Edinb]{M.~Bashkanov}
\author[IPJ,Kitz]{S.~D.~Bass}    
\author[ASWarsH]{M.~Ber{\l}owski}
\author[Budker,Novosib]{A.~Bondar}
\author[PGI,DUS]{M.~B\"uscher}
\author[IKPUU]{H.~Cal\'{e}n}  
\author[IFJ]{I.~Ciepa{\l}}   
\author[PITue,Kepler]{H.~Clement}    
\author[IPJ]{E.~Czerwi{\'n}ski}
\author[IKPJ]{R.~Engels}      
\author[ZELJ]{A.~Erven}
\author[ZELJ]{W.~Erven}
\author[Erl]{W.~Eyrich}    
\author[IKPJ,ITEP]{P.~Fedorets}   
\author[Giess]{K.~F\"ohl}  
\author[IKPUU]{K.~Fransson} 
\author[IKPJ]{F.~Goldenbaum} 
\author[IKPJ,IITI]{A.~Goswami} 
\author[IKPJ,HepGat]{K.~Grigoryev}
\author[IKPUU]{L.~Heijkenskj\"old \fnref{Mainz}}
\author[IKPJ]{V.~Hejny}
\author[NWU]{S.~Hirenzaki}  
\author[IPJ]{L.~Jarczyk}
\author[IKPUU]{T.~Johansson}  
\author[IPJ]{B.~Kamys}
\author[ANDES]{N.~G.~Kelkar}   
\author[ZELJ]{G.~Kemmerling \fnref{JCNS}}
%\author{G.~Khatri}\altaffiliation[present address: ]{\Harvard}\affiliation{\IPJ}
\author[IPJ]{A.~Khreptak}   
\author[HeJINR]{D. A.~Kirillov}  
\author[IPJ]{S.~Kistryn}    
\author[ZELJ]{H.~Kleines \fnref{JCNS}}
\author[Katow]{B.~K{\l}os}   
\author[NCBJ]{W.~Krzemie{\'n}}
\author[IFJ]{P.~Kulessa}
\author[IKPUU,ASWarsH]{A.~Kup\'{s}\'{c}}
%\author{A.~Kuzmin}      \affiliation{\Budker}\affiliation{\Novosib}
\author[NITJ]{K.~Lalwani}   
\author[IKPJ]{D.~Lersch \fnref{Florida}}      
\author[IKPJ]{B.~Lorentz}  
\author[IPJ]{A.~Magiera}     
\author[IKPJ,JARA]{R.~Maier}       
\author[IKPUU]{P.~Marciniewski}
\author[ASWarsN]{B.~Maria{\'n}ski}
\author[ASWarsN]{H.--P.~Morsch} 
\author[IPJ]{P.~Moskal}     
\author[IKPJ]{H.~Ohm}         
\author[IFJ]{W.~Parol}
\author[PITue,Kepler]{E.~Perez del Rio\fnref{INFN}}
\author[HeJINR]{N.M.~Piskunov}  
\author[IKPJ]{D.~Prasuhn}     
\author[IKPUU,ASWarsH]{D.~Pszczel}     
\author[IFJ]{K.~Pysz}        
%\author{A.~Pyszniak}    \affiliation{\IKPUU}\affiliation{\IPJ}
\author[IKPJ,JARA,Bochum]{J.~Ritman}   
\author[IITI]{A.~Roy}         
%\author{Z.~Rudy}        \affiliation{\IPJ}
\author[IPJ]{O.~Rundel}      
\author[IITB]{S.~Sawant}      
\author[IKPJ]{S.~Schadmand}   
\author[IPJ]{I.~Sch\"atti--Ozerianska}
\author[IKPJ]{T.~Sefzick}     
\author[IKPJ]{V.~Serdyuk}     
\author[Budker,Novosib]{B.~Shwartz}     
\author[PITue,Kepler,Tomsk]{T.~Skorodko}
\author[IPJ]{M.~Skurzok\fnref{INFN}}     
\author[IPJ]{J.~Smyrski}     
\author[ITEP]{V.~Sopov}      
\author[IKPJ]{R.~Stassen}     
\author[ASWarsH]{J.~Stepaniak}  
\author[Katow]{E.~Stephan}    
\author[IKPJ]{G.~Sterzenbach} 
\author[IKPJ]{H.~Stockhorst}  
\author[IKPJ,JARA]{H.~Str\"oher}   
\author[IFJ]{A.~Szczurek}    
\author[ASWarsN]{A.~Trzci{\'n}ski\fnref{dec}}
\author[IKPUU]{M.~Wolke}      
\author[IPJ]{A.~Wro{\'n}ska}
\author[ZELJ]{P.~W\"ustner}  
\author[KEK]{A.~Yamamoto} 
\author[ASLodz]{J.~Zabierowski}
\author[IPJ]{M. J.~Zieli{\'n}ski}
\author[IKPUU]{J.~Z{\l}oma{\'n}czuk}
\author[ASWarsN]{P.~{\.Z}upra{\'n}ski}
\author[IKPJ]{M.~{\.Z}urek\fnref{Berkeley}}

\address[IKPUU]{Division of Nuclear Physics, Department of Physics and Astronomy, Uppsala University, Box 516, 75120 Uppsala, Sweden}

\address[ASWarsN]{Department of Nuclear Physics, National Centre for Nuclear Research, ul.\ Pasteura 7, 02-093, Warsaw, Poland}

\address[IPJ]{Institute of Physics, Jagiellonian University, prof.\ Stanis{\l}awa {\L}ojasiewicza~11, 30-348 Krak\'{o}w, Poland}
 
\address[Edinb]{School of Physics and Astronomy, University of Edinburgh, James Clerk Maxwell Building, Peter Guthrie Tait Road, Edinburgh EH9 3FD, Great Britain}  
 
\address[Kitz]{Kitzb\"uhel Centre for Physics, Kitzb\"uhel, Austria}
 
\address[ASWarsH]{High Energy Physics Department, National Centre for Nuclear Research, ul.\ Pasteura 7, 02-093, Warsaw, Poland}
 
\address[Budker]{Budker Institute of Nuclear Physics of SB RAS, 11~akademika Lavrentieva prospect, Novosibirsk, 630090, Russia} 
 
\address[Novosib]{Novosibirsk State University, 2~Pirogova Str., Novosibirsk, 630090, Russia} 
 
\address[PGI]{Peter Gr\"unberg Institut, PGI--6 Elektronische Eigenschaften, Forschungszentrum J\"ulich, 52425 J\"ulich, Germany}

\address[DUS]{Institut f\"ur Laser-- und Plasmaphysik, Heinrich--Heine Universit\"at D\"usseldorf, Universit\"atsstr.~1, 40225 Düsseldorf, Germany}
 
\address[IFJ]{The Henryk Niewodnicza{\'n}ski Institute of Nuclear Physics, Polish Academy of Sciences, 152~Radzikowskiego St, 31-342 Krak\'{o}w, Poland} 

\address[PITue]{Physikalisches Institut, Eberhard--Karls--Universit\"at T\"ubingen, Auf der Morgenstelle~14, 72076 T\"ubingen, Germany} 
 
\address[Kepler]{Kepler Center f\"ur Astro-- und Teilchenphysik,
 Physikalisches Institut der Universit\"at T\"ubingen, Auf der 
 Morgenstelle~14, 72076 T\"ubingen, Germany} 
 
\address[IKPJ]{Institut f\"ur Kernphysik, Forschungszentrum J\"ulich, 52425 J\"ulich, Germany}

\address[ZELJ]{Zentralinstitut f\"ur Engineering, Elektronik und Analytik, Forschungszentrum J\"ulich, 52425 J\"ulich, Germany}

\address[Erl]{Physikalisches Institut, 
 Friedrich--Alexander--Universit\"at Erlangen--N\"urnberg, 
 Erwin--Rommel-Str.~1, 91058 Erlangen, Germany}

\address[ITEP]{Institute for Theoretical and Experimental Physics named by A.I.\ Alikhanov of National Research Centre ``Kurchatov Institute'', 25~Bolshaya Cheremushkinskaya, Moscow, 117218, Russia}

\address[Giess]{II.\ Physikalisches Institut, 
 Justus--Liebig--Universit\"at Gie{\ss}en, Heinrich--Buff--Ring~16, 35392 Giessen, Germany}
 
\address[IITI]{Department of Physics, Indian Institute of Technology Indore, Khandwa Road, Simrol, Indore--453552, Madhya Pradesh, India}
 
\address[HepGat]{High Energy Physics Division, Petersburg Nuclear Physics Institute named by B.P.\ Konstantinov of National Research Centre ``Kurchatov Institute'', 1~mkr.\ Orlova roshcha, Leningradskaya Oblast, Gatchina, 188300, Russia}

\address[NWU]{Department of Physics, Nara Women's University, Nara 630-8506, Japan}

\address[ANDES]{Departamento de Fisica, Universidad de los Andes, Cra.~1E, 18A--10, Bogot{\'a}, Colombia}  

\address[HeJINR]{Veksler and Baldin Laboratory of High Energiy Physics, Joint Institute for Nuclear Physics, 6~Joliot--Curie, Dubna, 141980, Russia}

\address[Katow]{August Che{\l}kowski Institute of Physics, University of Silesia, Uniwersytecka~4, 40-007, Katowice, Poland}

\address[NCBJ]{High Energy Physics Division, National Centre for Nuclear Research, 05-400 Otwock-Świerk, Poland}

\address[NITJ]{Department of Physics, Malaviya National Institute of Technology Jaipur, JLN Marg Jaipur - 302017, Rajasthan, India}

\address[JARA]{JARA--FAME, J\"ulich Aachen Research Alliance, 
 Forschungszentrum J\"ulich, 52425 J\"ulich, and RWTH Aachen, 52056 Aachen, Germany}
 
\address[Bochum]{Institut f\"ur Experimentalphysik I, Ruhr--Universit\"at Bochum, Universit\"atsstr.~150, 44780 Bochum, Germany}

\address[IITB]{Department of Physics, Indian Institute of Technology Bombay, Powai, Mumbai--400076, Maharashtra, India}

\address[Tomsk]{Department of Physics, Tomsk State University, 36~Lenina Avenue, Tomsk, 634050, Russia}

\address[KEK]{High Energy Accelerator Research Organisation KEK, Tsukuba, Ibaraki 305--0801, Japan} 

\address[ASLodz]{Department of Astrophysics, National Centre for Nuclear Research, 90--950 {\L}\'{o}d\'{z}, Poland}

%\address[Harvard]{Department of Physics, Harvard University, 17~Oxford St., Cambridge, MA~02138, USA}
 
\fntext[Mainz]{present address: Institut f\"ur Kernphysik, Johannes 
 Guten\-berg--Universit\"at Mainz, Johann--Joachim--Becher Weg~45, 55128 Mainz, 
 Germany}
 
 \fntext[JCNS]{present address: J\"ulich Centre for Neutron Science JCNS, Forschungszentrum J\"ulich, 52425 J\"ulich, Germany}
 
 \fntext[Florida]{present address: Department of Physics, Florida State University, 77 Chieftan Way, Tallahassee, FL 32306-4350, USA} 
 
\fntext[INFN]{present address: INFN, Laboratori Nazionali di Frascati, Via E. Fermi, 40, 
 00044 Frascati (Roma), Italy} 
 
\fntext[dec]{deceased} 

\fntext[Berkeley]{present address: Lawrence Berkeley National Laboratory, Berkeley, California 94720}

\begin{abstract}
We report on the experimental search for the bound state of an $\eta$ meson and $^{3}\hspace{-0.03cm}\mbox{He}$ nucleus performed using the WASA-at-COSY detector setup.
 %at COSY accelerator in Research Center J\"ulich in Germany with 
 %
 In order to search for the $\eta$-mesic nucleus decay, the \mbox{$pd\rightarrow$ $^{3}\hspace{-0.03cm}\mbox{He} 2\gamma$} and  $pd\rightarrow$ $^{3}\hspace{-0.03cm}\mbox{He} 6\gamma$ channels have been analysed.
 These reactions manifest the direct decay of the $\eta$ meson bound in a $^{3}\hspace{-0.03cm}\mbox{He}$ nucleus.~This non-mesonic decay channel has been considered for the first time. When taking into account only statistical errors, the obtained excitation functions reveal a slight indication for a possible bound state signal corresponding to %\textit{\textcolor{blue}{MS:
 a $^3$He-$\eta$ nucleus width $\Gamma$ above 20 MeV and binding energy $B_s$ between 0 and 15 MeV.
 %}}
  However, the determined cross sections are consistent with zero in the range of the systematic uncertainty. Therefore, as final result we estimate only the upper limit for the cross section of the $\eta$-mesic $^{3}\hspace{-0.03cm}\mbox{He}$ nucleus formation followed by the $\eta$ meson decay which varies between $2$~nb and $15$~nb depending on possible bound state parameters.
\end{abstract}

%\pacs{13.60.Le, 14.40.Aq}    % PACS, the Physics and Astronomy
                             % Classification Scheme.
%


\end{frontmatter}

\section{Introduction}

Strong attractive interactions between the $\eta$ meson and nucleons mean that there is a chance to form $\eta$ meson bound states in nuclei \cite{Haider:1986sa}.
If discovered in experiments, these mesic nuclei would be a new state of matter bound just by the strong interaction without electromagnetic Coulomb effects playing a role. Strong interaction bound states are formed in a different way as compared to exotic atoms which involve binding of electrically charged mesons with nuclei.
For the latter, negatively charged pions or kaons could replace an electron in an outer orbital in a standard atom and get bound in the atom due to the Coulomb interaction.~The charged meson in such an excited state quickly undergoes transitions to the lower states until it is close enough to the nucleus and is either absorbed by the nucleus or lost in a nuclear reaction.
For strong interactions,
in contrast to the pion, the neutral $\eta$ meson is special due to the strong attractive nature of this meson-nucleon interaction~\cite{Haider:1986sa}.~An off-shell $\eta$ meson produced in nuclear reactions such as the $pd\rightarrow$ $^{3}\hspace{-0.03cm}\mbox{He} 2\gamma$ and $pd\rightarrow$ $^{3}\hspace{-0.03cm}\mbox{He} 6\gamma$ below the $\eta$ production threshold may form a bound state with the nucleus within which it is produced. Thus the absence of the electromagnetic interaction and the attractive nature of the $\eta$-nucleon interaction, makes the case of the neutral $\eta$ meson different from that of the pion or the kaon and opens the possibility for an exotic nucleus made up of the meson and nucleons.
%Strong attractive interactions between the $\eta$ meson and nucleons mean that there is a chance to form $\eta$ meson bound states in nuclei \cite{Haider:1986sa}.
%If discovered in experiments, these mesic nuclei would be a new state of matter bound just by the strong interaction without electromagnetic Coulomb effects playing a role. 
Early experiments with low statistics
using photon~\cite{Pheron:2012aj,Baskov:2012yd}, 
pion~\cite{Chrien:1988gn}, proton~\cite{Budzanowski:2008fr} 
or deuteron~\cite{Afanasiev:2011zza,Moskal:2010ee,Adlarson2013,Adlarson:2016dme}
beams gave hints for possible $\eta$ mesic bound states but no clear signal \cite{Metag:2017yuh,Kelkar:2013lwa}.
%More recent COSY searches have focused on possible $\eta$ 
%bound states in $^3$He and $^4$He \cite{Adlarson:2013xg,Adlarson:2016dme}
%**with limits from Helium-4**.
%

Here we present a new high statistics search for \mbox{$^3$He-$\eta$} bound states with data from the WASA-at-COSY experiment. We focus on the two main neutral decay channels of the $\eta$ meson: 
$\eta \to 2\gamma$ with branching ratio
39.41$\pm$0.20\% and $\eta \to 3\pi^0 \to 6\gamma$ 
with branching ratio 
%\textcolor{blue}{MS: $32.68 \pm 0.23 \%$
%** SB: Please check
%this again.
%The 32.68 number is for 
%$\eta \to 3 \pi^0$.
%But there is also a 
%BR of $98.823 \pm 0.034$
%\% for each of the 3
%$\pi^0 \to 2 \gamma$
%decays. My combining of
%these gave my estimate
%of 
$31.54 \pm 0.22$\%~\cite{PDG}.
These processes constitute more than 70\% of the $\eta$ decays. 
The choice of neutral decay channels
minimizes final state interactions
involving charged particles.
Concurrent measurement of the two channels increases the statistics and enables one to control systematic uncertainties in photons detection.
The two-photon decay was previously suggested
in \cite{Bass:2005hn}
as a clean probe of the $\eta$ in nuclear media. 
%assuming that one knows whether the $\eta$ is formed at the centre or edge of the nucleus.

Considering the $\eta$-nucleus interaction, bound states can be formed by the attractive interaction with finite level width corresponding to the finite lifetime of the state due to the absorptive interaction with the nucleus. The momentum distribution of the bound $\eta$ meson determines the sum of the momenta of the emitted photons.
%and the 
Nuclear absorption and the additional $\eta$ decay (disappearance) processes, reduces significantly the in-medium branching ratio of 2$\gamma$ and 6$\gamma$ decay channels~\cite{Skurzok_Hirenzaki2019}.

%\textcolor{red}{There exist several models for the $\eta$-nucleus interaction in literature. These in turn are based on models of the $\eta$-nucleon interaction, the strength of which has still remained to be a matter of debate.~The latter is what one would indeed like to pin down from experiment. The task would have been easier of course if the existence of an $\eta$ mesic bound state as predicted by one of the models would have been found. In the absence of such a finding, the data for this reaction is still useful as future theoretical calculations of these reactions could narrow the uncertainty in the understanding of the $\eta$-nucleon (and hence $\eta$-nucleus) interaction.}

$\eta$ meson interactions with nucleons and nuclei are a topic 
of great experimental and 
theoretical interest.
For recent reviews see 
\cite{Metag:2017yuh,Kelkar:2013lwa,Bass:2018xmz,Krusche:2014ava,Wilkin:2016mfn}.
%\textcolor{red}{There exist several models for the $\eta$-nucleus interaction in literature. These in turn are based on models of the $\eta$-nucleon interaction, the strength of which has still remained to be a matter of debate.}
Possible $\eta$-nucleus binding energies are related
to the $\eta$-nucleon optical potential and to the
value of $\eta$-nucleon scattering length $a_{\eta N}$
\cite{Ericson:1988gk}. 
Phenomenological 
estimates for the real part 
of $a_{\eta N}$ are typically
between 0.2 and 1 fm
depending on the model assumptions.
$\eta$ bound states in helium require a large $\eta$-nucleon 
scattering length 
with real part greater than about 0.7--1.1~fm  
\cite{Barnea:2017epo,Barnea:2017oyk,Fix:2017ani}.
Recent calculations in the framework of 
optical potential~\cite{Xie:2016zhs},
multi-body calculations~\cite{Barnea:2017oyk},
and pionless effective field theory~\cite{Barnea:2017epo}
suggest a possible $^3$He-$\eta$ bound state.

%\textcolor{red}{The $\eta$-nucleon interaction is what one would like to pin down from experiment. The task would have been easier of course if the existence of an $\eta$ mesic bound state as predicted by one of the models would have been found. In the absence of such a finding, our experimental data is still useful as future theoretical calculations of these reactions could narrow the uncertainty in the understanding of the $\eta$-nucleon (and hence $\eta$-nucleus) interaction.}

Modifications of meson properties
are expected in medium.~In studies of the transparency of nuclei to propagating mesons produced in photoproduction experiments one finds strong $\eta$ absorption in nuclei~\cite{Nanova:2012vw}. 
%\textcolor{red}{N. K. to remove: (to good approximation, the $\eta$ meson is emitted only from the nuclear surface)}
For the $\eta'$ one finds weaker interaction with the nucleus. An effective mass shift for the $\eta'$ in medium has been
observed by the CBELSA/TAPS Collaboration~\cite{Nanova:2013fxl}.
The $\eta'$-nucleus optical potential 
$V_{\rm opt} = V_{\rm real} + iW$
deduced from these photoproduction experiments with a 
carbon target is 
%\begin{eqnarray}
$V_{\rm real} (\rho_0)
= m^* - m 
= -37 \pm 10 \pm 10 \ {\rm MeV}
$
and
$
W(\rho_0) = -10 \pm 2.5 \ {\rm MeV}$
at nuclear matter density $\rho_0$.
This mass shift is very close to the prediction of the
Quark Meson Coupling mode (QMC)
with mixing angle -20 degrees \cite{Bass:2005hn,Bass:2013nya},
which also predicts a potential depth about -100 MeV 
for the $\eta$ at $\rho_0$.
The $\eta'$ results are also
consistent with scattering length estimates from COSY-11~\cite{Czerwinski:2014yot} and Bonn~\cite{Anisovich:2018yoo}. Experimental search for $\eta'$ - nucleus bound states has also been performed with results reported in Ref. \cite{Tanaka}.

Hints for possible $\eta$ helium bound states are inferred from the observation of strong interaction in the $\eta$ helium system. One finds a sharp rise in the cross section 
at threshold for $\eta$ production in both
photoproduction from $^3$He 
\cite{Pheron:2012aj,Pfeiffer:2003zd}
and in the proton-deuteron reaction
$dp\rightarrow$ $^{3}\hspace{-0.03cm}\mbox{He} \eta$
\cite{Adlarson:2018rgs}.
These observations may hint at a reduced $\eta$ effective mass in the nuclear medium.

%$pd\rightarrow$ $^{3}\hspace{-0.03cm}\mbox{He} 2\gamma$

Previous bound state searches at COSY have been focused on the reaction
$dd\rightarrow$ $^{3}\hspace{-0.03cm}\mbox{He} N \pi$~\cite{Adlarson2013,Adlarson:2016dme}. Studies of the excitation function around the threshold for $dd\rightarrow$ $^{4}\hspace{-0.03cm}\mbox{He} \eta$
did not reveal a structure 
that could be interpreted as a narrow mesic nucleus. 
Upper limits for the total cross sections for bound state production and decay in the processes
$dd\rightarrow$ ($^{4}\hspace{-0.03cm}\mbox{He}$-$\eta)_{bound} \rightarrow$ $^{3}\hspace{-0.03cm}\mbox{He} n \pi^{0}$ 
and 
\mbox{$dd\rightarrow$ ($^{4}\hspace{-0.03cm}\mbox{He}$-$\eta)_{bound} \rightarrow$ $^{3}\hspace{-0.03cm}\mbox{He} p \pi^{-}$}
were deduced to be about 5~nb and 10~nb for 
the $n \pi^0$ and $p \pi^-$ channels respectively \cite{Adlarson:2016dme}.
The bound state production 
cross sections 
for $pd\rightarrow$ ($^{3}\hspace{-0.03cm}\mbox{He}$-$\eta)_{bound}$ \cite{Wilkin:2014mla} 
are expected to be more than 20 times larger than for
$dd\rightarrow$ ($^{4}\hspace{-0.03cm}\mbox{He}$-$\eta)_{bound}$~\cite{Wycech:2014wua}.

In May 2014 the experiment searching for $\eta$ mesic $^3$He nuclei was performed at the COSY accelerator~\cite{COSY_description1,COSY_description5} in J\"ulich, Germany.
The measurements were carried out using the WASA-at-COSY detector~\cite{WASA_description1,WASA_description2,WASA_description3,WASA_description4,3Heeta_cross_section}.
The mesic nuclei are supposed to be formed in proton-deuteron collisions.
A ramped proton beam with beam momentum varying in the range from $1.426$ to $1.635$~GeV/c corresponding to $^3$He$\eta$ excess energy range from $-70$ to $30$~MeV and a pellet deuterium target~\cite{WASA_pellett_target} were used.
The $^3$He-$\eta$ bound state was searched for in the $pd\rightarrow$ ($^{3}\hspace{-0.03cm}\mbox{He}$-$\eta)_{bound} \rightarrow$ $ ^{3}\hspace{-0.03cm}\mbox{He}2\gamma$  and $pd\rightarrow$ ($^{3}\hspace{-0.03cm}\mbox{He}$-$\eta)_{bound} \rightarrow$ $^{3}\hspace{-0.03cm}\mbox{He}6\gamma$ decay channels. These channels that manifest the direct decay of $\eta$ bound in $^3$He nucleus have been investigated for the first time. The existence of the bound $^3$He-$\eta$ state would manifest itself as a maximum or interference pattern in the excitation function for both of the studied reactions below the $pd\rightarrow ^3$He$\eta$ reaction threshold.

%In order to determine the absolute values for the excitation function, the luminosity was established depending on the excess energy.

For the normalization of the excitation functions, the integrated luminosity was determined as a function of the excess energy.~The analysis is presented in the next section. Further on, the data selection and efficiency determination is described. The data analysis is followed by the interpretation of the achieved excitation functions in view of the possible signal from the $\eta$-mesic $^3$He. 
 
%\textit{\textcolor{green}{PM:
%I think we may skip the below part ;
%{\it
%At end, how many sigma is the %statistical
%result before systematics ?
%* PM ****
%The statistical significance of the signal is large.
%Probably more than 5 sigma.  But I am not sure if we should mention this ?

\section{Luminosity determination}
\begin{figure}[h!]
\centering
	\includegraphics[width=13.0cm,height=6.0cm]{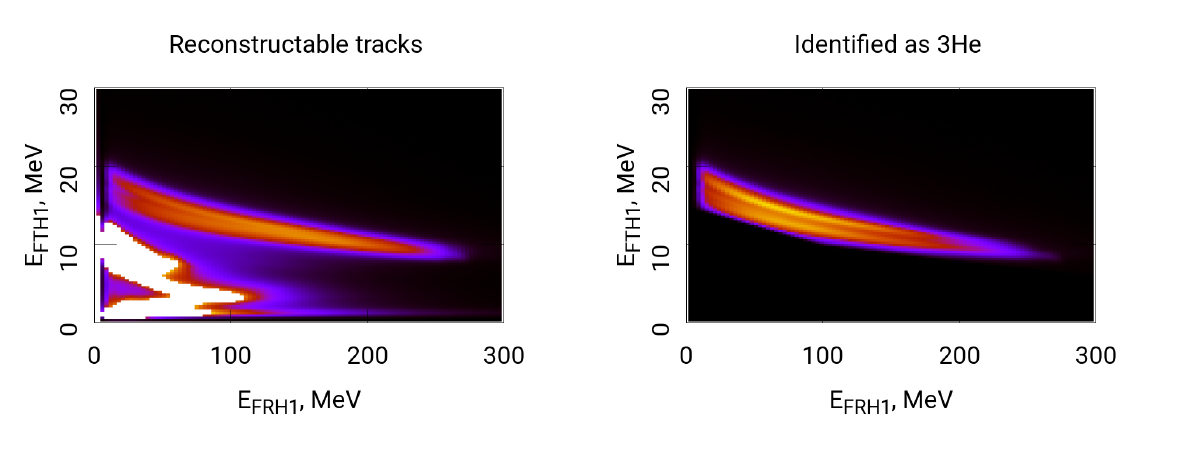}
	\caption{2-D histograms of energies deposited in the first layer of Forward Trigger Hodoscope (FTH1) and the first layer of Forward Range Hodoscope (FRH1) for all events with signal in Forward Proportional Chamber (FPC) (left panel) and events that were identified as $^{3}\hspace{-0.03cm}\mbox{He}$ (right panel).}\label{fig_pid}
\end{figure}

Luminosity was determined based on the $pd\rightarrow ^3$He$\eta$ and $pd\rightarrow ppn_{spectator}$ reactions.
The $pd\rightarrow ^3$He$\eta$ reaction analysis allows one to estimate the integrated luminosity for $^3$He$\eta$ excess energy $Q_{^3\hspace{-0.05cm}He\eta}$ above zero. The $^3$He particles were registered in the forward detector~\cite{WASA_description1} and identified using the $\Delta E-E$ method based on energy losses in scintillator layers (see~Fig.~\ref{fig_pid}).

%Spectrum of energy deposited in the first layer of Forward Window Counter (FWC1) and the first layer of Forward Range Hodoscope (FRH1) for experimental data. The selected area for 3He is marked with black line. The empty area below comes from the preselection conditions.

\begin{figure}[h!]
	\includegraphics[width=10.0cm,height=6.0cm]{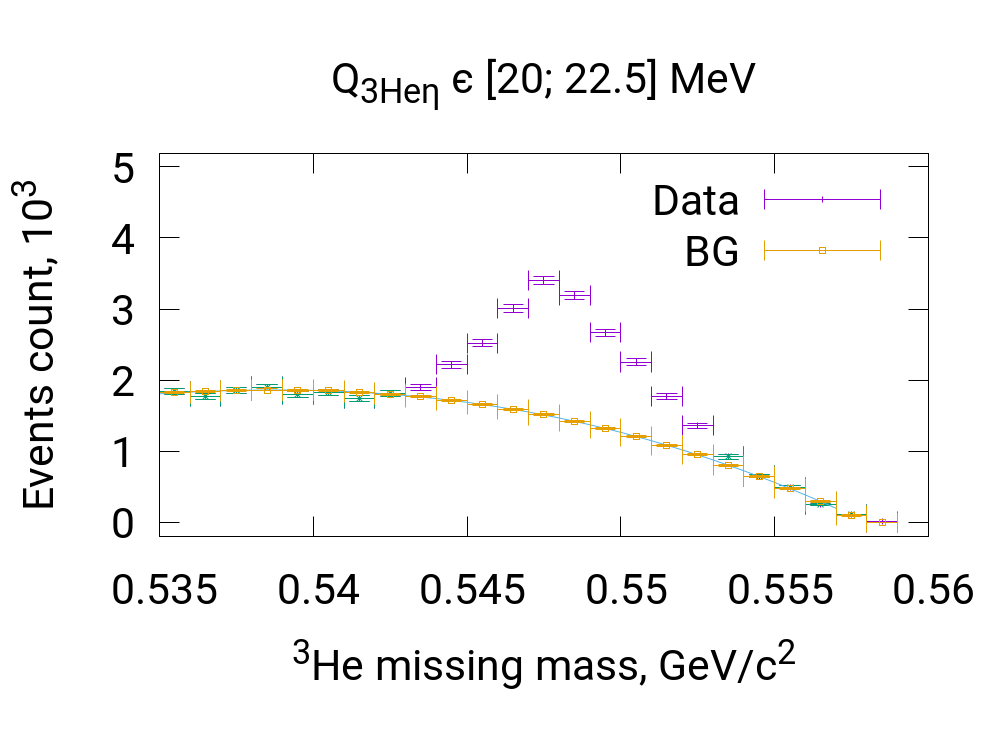}
	\caption{\small
		$^3$He missing mass spectrum obtained from data for the excess energy range of $Q_{^3\hspace{-0.05cm}He\eta}~\in~[20.0;22.5]$~MeV. The part of the spectrum that is considered to be background is shown with green color and is fitted with a polynomial of fourth power (orange).
	}
	\label{he3_mm_spectra_fit}
\end{figure}

The count of events originating from this reaction was obtained based on the $^3$He missing mass spectra for each excess energy interval separately. An example spectrum is shown in Fig.~\ref{he3_mm_spectra_fit}.
The reconstruction efficiency was calculated using Monte Carlo simulations taking into account the experimental data on cross sections and angular distributions~\cite{3Heeta_cross_section,3Heeta_cross_section_older1,3Heeta_cross_section_older2,3Heeta_cross_section_older3}.
%The analysis of this reaction also allowed to determine the beam momentum correction constant for the ramped beam from the kinematic conditions.

The $pd\rightarrow ppn_{spectator}$ reaction analysis allows one to determine the integrated luminosity for the whole beam momentum range.~As far as the target overlapping by the beam is changing during the acceleration cycle, the integrated luminosity value can change depending on the beam momentum.~The registration efficiency for the $pd\rightarrow ppn_{spectator}$ reaction was obtained with dedicated Monte Carlo simulations described  in Refs.~\cite{OleksandrPhD-arXiv,czyzyk}.
The distribution of relative proton-neutron motion inside the target deuteron was calculated based on the parametrisation of the Paris potential~\cite{PARIS_model}. Data on the proton-proton elastic scattering cross section and 
the angular distribution 
%from Ref.
~\cite{PP_CS_link} were used for simulating the quasi-elastic scattering in the framework of the spectator model.~The calculated cross section was multiplied by the factor $0.96$ 
to take into account the shading effect~\cite{shading_effect}. It is worth noting that above the $\eta$ production threshold, the two estimates of luminosity are in agreement (based on the $pd\rightarrow ppn_{spectator}$ and $pd\rightarrow ^3$He$\eta$ reactions~\cite{OleksandrPhD-arXiv}). The total integrated luminosity was determined to be $2446\pm3(stat.)\pm66(syst.)\pm4$(norm.)~nb$^{-1}$
where the statistical, systematic and normalisation errors are indicated, respectively~\cite{OleksandrPhD-arXiv}. This is the largest statistics ever obtained for these experimental conditions.

%%%%%%%%%%%%%%%%%%%%%%%%%%%%%%%%%%%%%%%%%%%%%%%%%%%%

\section{The analysis of $pd\rightarrow$ ($^{3}\hspace{-0.03cm}\mbox{He}$-$\eta)_{bound} \rightarrow$ $^{3}\hspace{-0.03cm}\mbox{He}2\gamma$ and $pd\rightarrow$ ($^{3}\hspace{-0.03cm}\mbox{He}$-$\eta)_{bound} \rightarrow$ $^{3}\hspace{-0.03cm}\mbox{He}6\gamma$ reactions}

As a first step, in order to establish the optimal selection criteria, Monte Carlo simulations for the $pd\rightarrow$ ($^{3}\hspace{-0.03cm}\mbox{He}$-$\eta)_{bound} \rightarrow$ $^{3}\hspace{-0.03cm}\mbox{He}2\gamma$ and $pd\rightarrow$ ($^{3}\hspace{-0.03cm}\mbox{He}$-$\eta)_{bound} \rightarrow$ $^{3}\hspace{-0.03cm}\mbox{He}6\gamma$ reactions were performed in the framework
of the spectator model with the assumption of an isotropic distribution of bound $\eta$ meson decay products 
%emission 
in its rest frame. The momentum of the $\eta$ meson was simulated using the recent
%assuming recently elaborated 
model~\cite{Skurzok_Hirenzaki2019} in which the $^3$He-$\eta$ relative momentum distribution was calculated by solving the Klein-Gordon equation assuming the potential of $\eta$-nucleus interaction based on Hiyama's density distribution in $^3$He~\cite{Hiyama1,Hiyama2,Hiyama3}.

\begin{figure}[h!]
	\includegraphics[width=6.0cm,height=6.0cm]{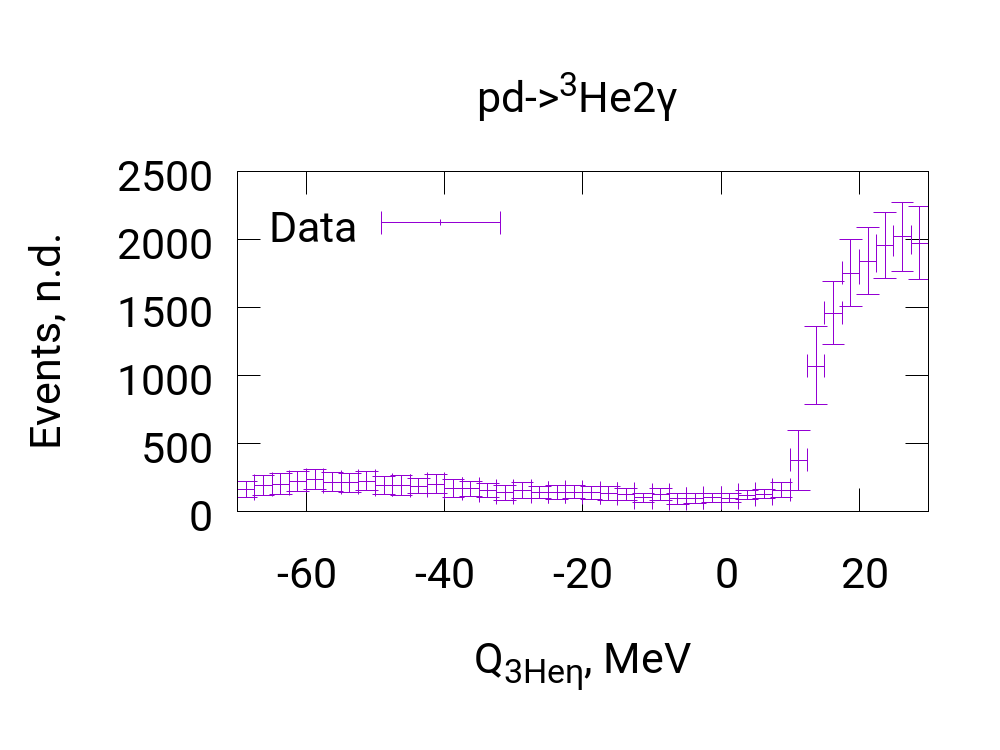}
	\includegraphics[width=6.0cm,height=6.0cm]{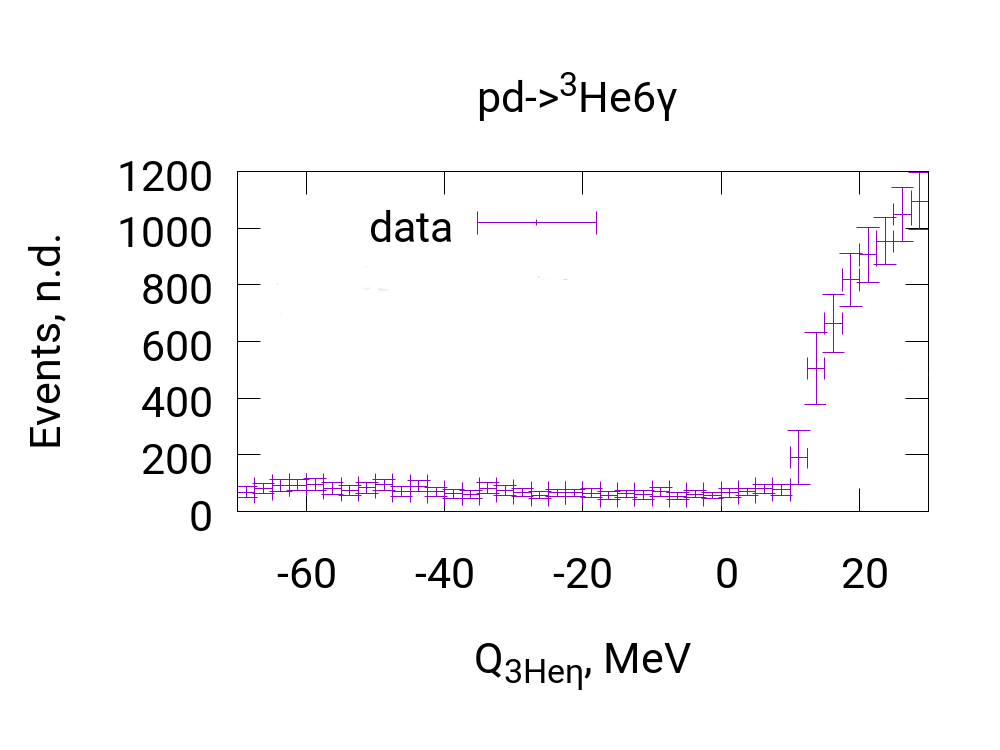}
\caption{The dependence of determined events count on $Q_{^3\hspace{-0.05cm}He\eta}$ for $pd\rightarrow$ $^{3}\hspace{-0.03cm}\mbox{He}2\gamma$ reaction (left panel) and $pd\rightarrow$ $^{3}\hspace{-0.03cm}\mbox{He}6\gamma$ reaction (right panel). The error bars include both statistical and systematic uncertainties.}
	\label{two_gamma_events}
\end{figure}

For the $pd\rightarrow$ ($^{3}\hspace{-0.03cm}\mbox{He}$-$\eta)_{bound} \rightarrow$ $^{3}\hspace{-0.03cm}\mbox{He}2\gamma$ reaction analysis, the events containing a $^3$He track in the forward detector and at least two photons in the central detector were selected.
If there were more than two photons, the pair with the invariant mass closest to the $\eta$ mass corrected by $Q_{^3\hspace{-0.05cm}He\eta}$ value was chosen.
Then the restrictions on $^3$He missing mass, $\gamma$-$\gamma$ missing mass, and $\gamma$-$\gamma$ invariant mass were applied using selection ranges based on the simulated distributions~\cite{OleksandrPhD-arXiv}.
The excitation function obtained for the $pd\rightarrow$ $^{3}\hspace{-0.03cm}\mbox{He}2\gamma$ reaction is shown in the left panel of Fig.~\ref{two_gamma_events}. 
%The total events count for $Q_{^3He\eta}>0$ is in agreement with the calculations taking into account the $pd\rightarrow^3$$He\eta$ reaction.

\begin{figure}[h!]
%\centering
	\includegraphics[width=6.0cm,height=6.0cm]{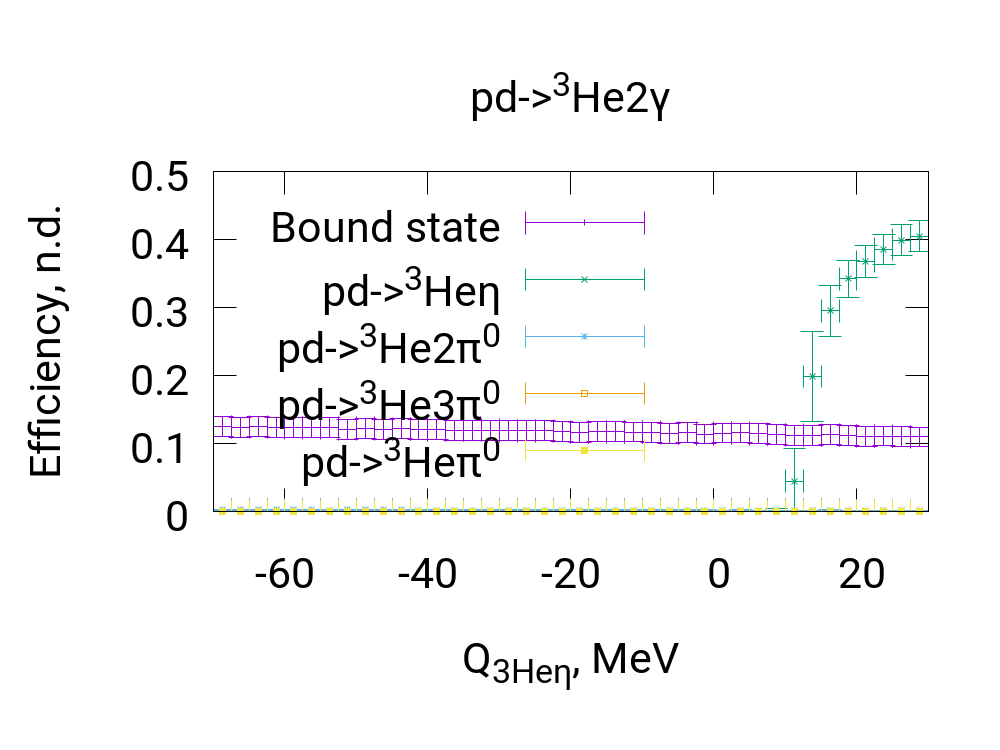}
	\includegraphics[width=6.0cm,height=6.0cm]{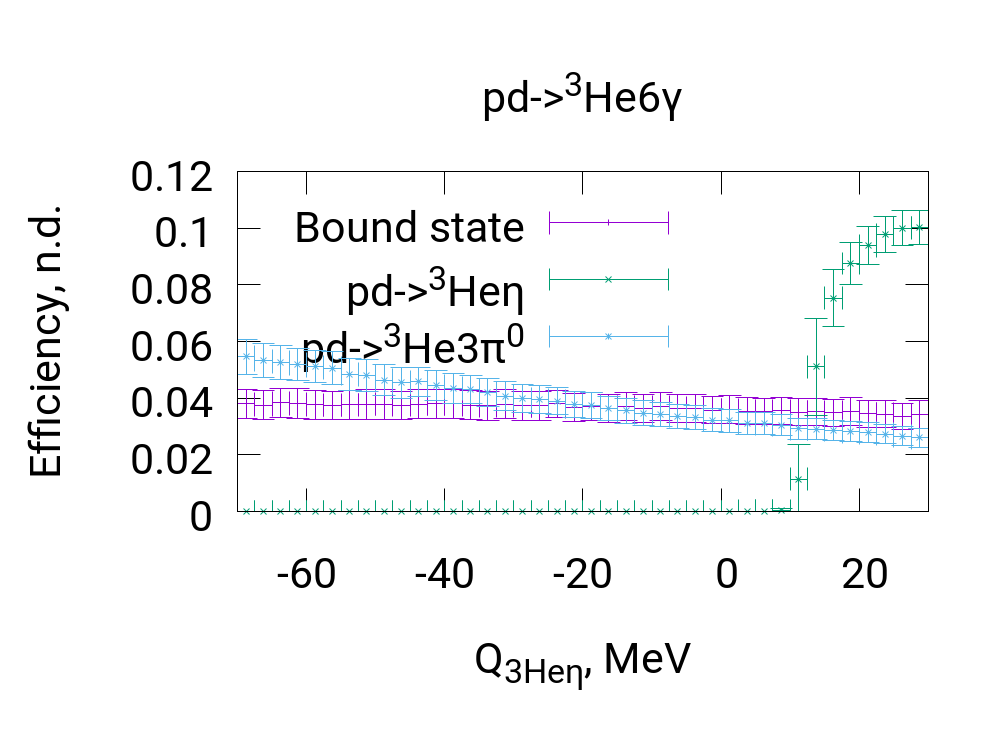}
	\caption{The efficiency for different reactions when applying selection criteria defined for the $pd\rightarrow$ $^{3}\hspace{-0.03cm}\mbox{He} 2\gamma$ (left) and $pd\rightarrow$ $^{3}\hspace{-0.03cm}\mbox{He} 6\gamma$ (right) reaction analysis.}\label{effic}
\end{figure}

The signal from the bound state is expected for excess energies around or below zero. The increase of events above 10~MeV is due to the $pd\rightarrow^3$He$\eta$ reaction. It starts at 10~MeV because of a hole for the COSY beam in the geometrical acceptance of the WASA-at-COSY detector (see Fig.~\ref{effic}). 

For the $pd\rightarrow$ ($^{3}\hspace{-0.03cm}\mbox{He}$-$\eta)_{bound} \rightarrow$ $^{3}\hspace{-0.03cm}\mbox{He}6\gamma$ reaction analysis, the events containing a $^3$He track in the forward detector and at least six photons in the central detector were selected.
For each combination forming three pairs, to identify the $\eta\rightarrow3\pi^0\rightarrow6\gamma$ decay, the following quantity is calculated:
\begin{equation}
	D= \sum_{i=1}^{3} (m_{\gamma_{(2i-1)}\gamma_{2i}} - m_{\pi^0})^2 
	\label{6gamma_3pi0_identification}
\end{equation}
where $m_{\gamma_{(2i-1)}\gamma_{2i}}$ is the $\gamma$ pair invariant mass and 
$m_{\pi^0}$ is $\pi^0$ mass. The combination of six photons
%$\gamma$s 
that minimises $D$ was choosen.
Then analogous to the $2\gamma$ case, the selection conditions on the $^3$He missing mass, $6\gamma$ invariant mass, and $6\gamma$ missing mass were applied based on the simulated distributions~\cite{OleksandrPhD-arXiv}. The excitation function obtained for the $pd\rightarrow$ $^{3}\hspace{-0.03cm}\mbox{He}6\gamma$ reaction is shown in the right panel of Fig.~\ref{two_gamma_events}.

\begin{figure}[h!]
		\includegraphics[width=10.0cm,height=6.0cm]{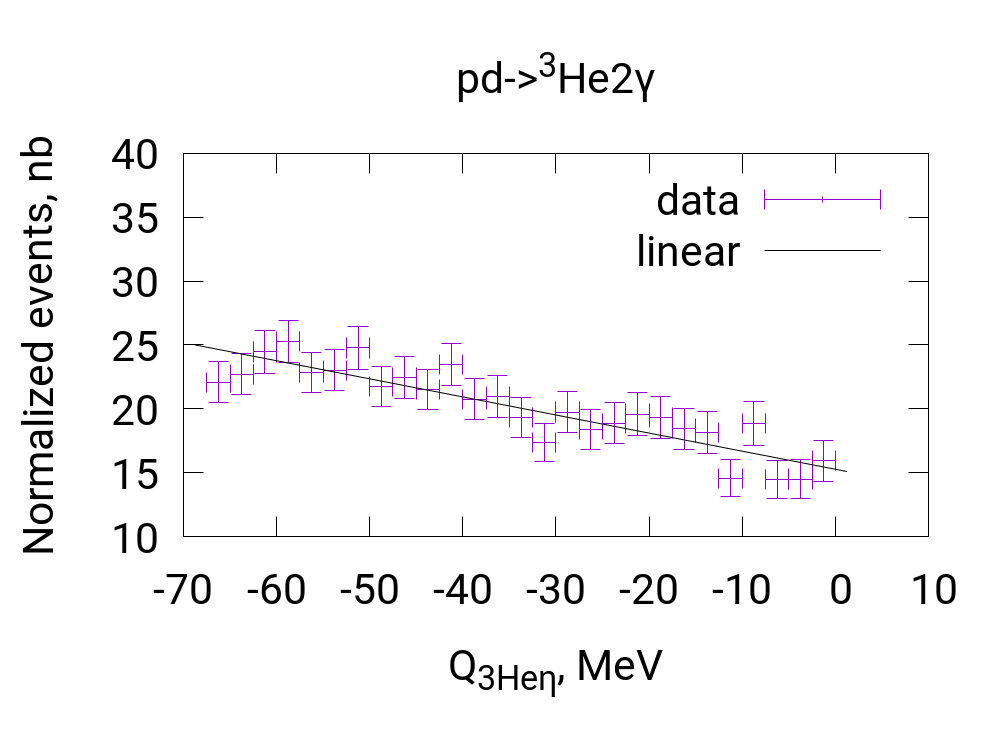}
		\includegraphics[width=10.0cm,height=6.0cm]{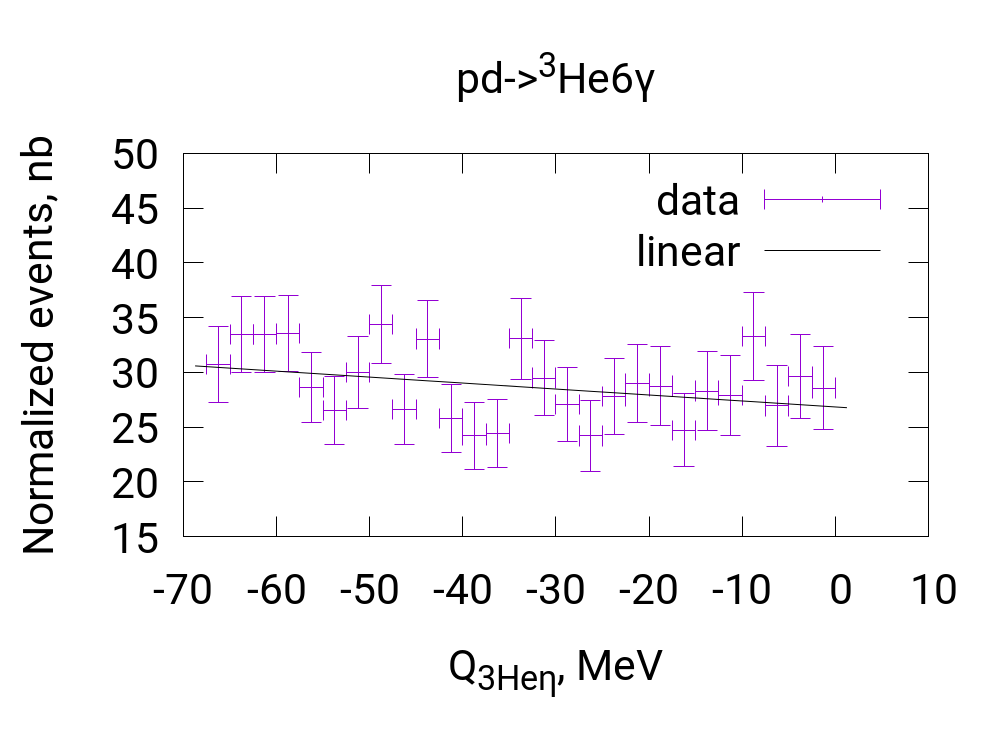}
	\caption{
		Excitation curves determined for the
		$pd\rightarrow$ ($^{3}\hspace{-0.03cm}\mbox{He}$-$\eta)_{bound} \rightarrow$ $^{3}\hspace{-0.03cm}\mbox{He}2\gamma$ (upper panel) and $pd\rightarrow$ ($^{3}\hspace{-0.03cm}\mbox{He}$-$\eta)_{bound} \rightarrow$ $^{3}\hspace{-0.03cm}\mbox{He}6\gamma$ (lower panel) reactions.
		Superimposed lines indicate result of the fit of the line.
		The points above the $\eta$ production threshold are excluded from the analysis.
	}
	\label{excitation_fit_linear}
\end{figure}

The excitation curves have been normalised using the integrated luminosity values calculated based on the $pd\rightarrow ppn_{spectator}$ reaction and the efficiency determined based on Monte Carlo simulations.
The results for both studied reactions are shown in Fig.~\ref{excitation_fit_linear}.

\section{The upper limit for the $\eta$ mesic \mbox{$^3$He} production cross section}

The excitation curves obtained in the analysis (Fig.~\ref{excitation_fit_linear}) did not reveal any resonance-like structures and the fit with linear functions results in $\chi^2$ value $<$ 1 when normalized to the number of degrees of freedom. 
This indicates that no strong signal from the bound $^3$He-$\eta$ state is observed.  

Further on, for the quantitative estimates of the upper limits for the bound state production, a fit to the excitation curves with a linear function (for background) plus  a Breit-Wigner function (for the signal) was performed. The fit was done for different combinations of the assumed $\eta$-mesic $^3$He binding energies $B_s$ and widths $\Gamma$. The value of $\Gamma$ was tested in the range from 1.25~MeV to 38.75~MeV (with the step of 2.5~MeV) and  $B_s$ in the range from 1.25~MeV to 63.75~MeV (with the step of 2.5~MeV).
%}}
\begin{figure}[h!]
		\includegraphics[width=10.0cm,height=6.0cm]{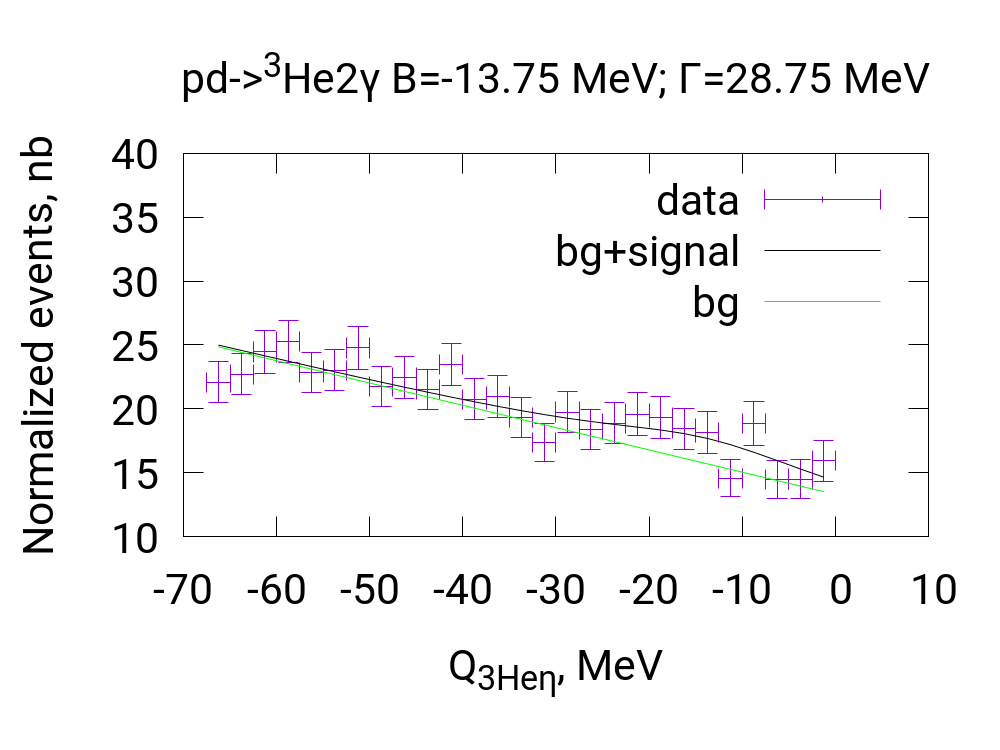}
		\includegraphics[width=10.0cm,height=6.0cm]{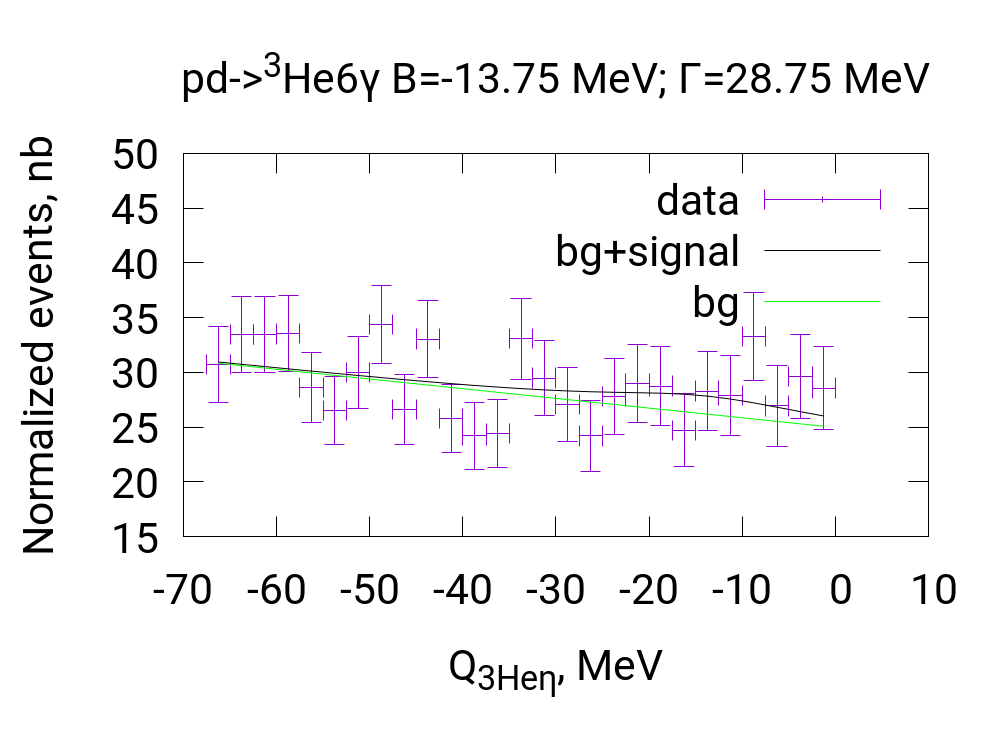}
	\caption{
		Exemplary result of the simultaneous fit of functions~\ref{excitation_curves_fit_function1} and \ref{excitation_curves_fit_function2} to the experimental data for the assumed $B_s$ and $\Gamma$ values as indicated above the figures. Superimposed 
		black line shows the full fit result, and the green line shows the background function only.
	}
	\label{excitation_fit_BW}
\end{figure}

\begin{figure}[h!]
		\includegraphics[width=10.0cm,height=6.0cm]{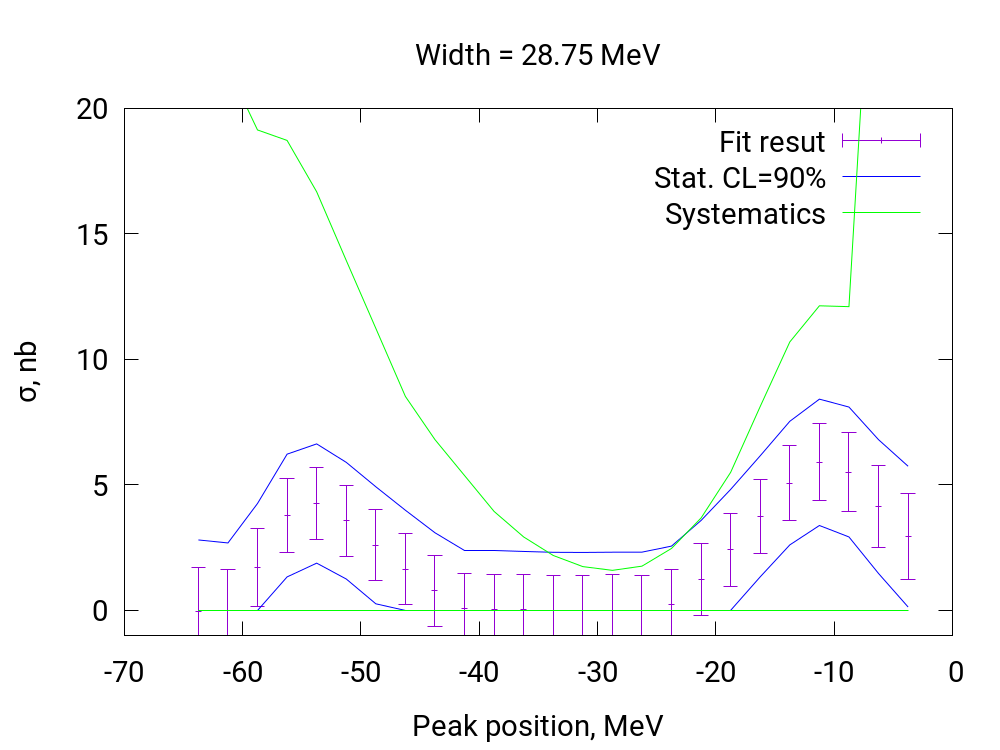}
	\caption{
		Upper limits for the bound state production cross section via $pd\rightarrow$ ($^{3}\hspace{-0.03cm}\mbox{He}$-$\eta)_{bound} \rightarrow$ $^{3}\hspace{-0.03cm}\mbox{He}$($\eta~ \mbox{decays})$ 
		as function of binding energy for fixed width $\Gamma$=28.75~MeV. 
		The values of the Breit-Wigner amplitude $\sigma$ are shown with statistical uncertainties. The range of possible bound state production cross section obtained based on statistical uncertainty corresponding to $90\%$ confidence level is shown by blue lines.
		The range of possible bound state production cross section including  systematic uncertainty is shown by green lines.
	}
	\label{upper_limit_1d}
\end{figure}

For a given $B_s$ and $\Gamma$ pair, the following functions were fit simultaneously for the two studied reaction channels:
\begin{equation}
	\rho^{fit}_{^3\hspace{-0.05cm}\mathrm{He}2\gamma}(Q_{^3\hspace{-0.05cm}\mathrm{He}\eta}) = P_{\eta\rightarrow2\gamma} \cdot \sigma \cdot \sigma_b(Q_{^3\hspace{-0.05cm}\mathrm{He}\eta}) + p_1 Q_{^3\hspace{-0.05cm}\mathrm{He}\eta} + p_2,
	\label{excitation_curves_fit_function1}
\end{equation}
\begin{equation}
\rho^{fit}_{^3\hspace{-0.05cm}\mathrm{He}6\gamma}(Q_{^3\hspace{-0.05cm}\mathrm{He}\eta}) = P_{\eta\rightarrow6\gamma} \cdot \sigma \cdot \sigma_b(Q_{^3\hspace{-0.05cm}\mathrm{He}\eta}) + p_3 Q_{^3\hspace{-0.05cm}\mathrm{He}\eta} + p_4.
\label{excitation_curves_fit_function2}
\end{equation}
Here $\sigma$, $p_1$, $p_2$, $p_3$, and $p_4$ are the free fit parameters, $P_{\eta\rightarrow2\gamma}$ and $P_{\eta\rightarrow6\gamma}$ are the branching ratios for the $\eta\rightarrow 2\gamma$
and $\eta\rightarrow 6\gamma$ decays. 
%(Fig~\ref{excitation_fit_linear}).
Assuming that the ratio of branching ratios for the $\eta\rightarrow2\gamma$ and $\eta\rightarrow3\pi^0$ decay channels for the bound $\eta$ meson remain the same as in vacuum, the vacuum branching ratio values of
	$P_{\eta\rightarrow2\gamma}$~=~0.3941 and 
	$P_{\eta\rightarrow3\pi^0\to 6\gamma}$~=~0.3268
were used for performing the fit~\cite{PDG}. The function $\sigma_b(Q_{^3\hspace{-0.05cm}He\eta})$ in the fit formulae represents a Breit-Wigner shape which for a given values of $B_s$ and $\Gamma$ reads:
\begin{equation}
	\sigma_b(Q_{^3He\eta},B_s,\Gamma) 
	= \sigma~\frac{\Gamma^2/4}{(Q_{^3\hspace{-0.05cm}He\eta}-B_s)^2+\Gamma^2/4}.
	\label{Breight_Wigner_formulae}
\end{equation}
%
%\textit{\textcolor{green}{PM:
Example results of the fit are shown in Fig.~\ref{excitation_fit_BW}. The figure shows results for the $B_s$ and $\Gamma$ values (indicated above the plots) for which the fitted values of $\sigma$ differ from zero with the largest statistical significance.  
Fig.~\ref{upper_limit_1d} indicates the results of the fit as a function of the $B_s$ for the most promising value of $\Gamma~=~28.75$~MeV. 

The upper limit of the total cross section was determined based on the fit parameter uncertainty $\Delta\sigma^{stat}$:

\begin{equation}
\sigma^{CL=90\%}_{upper}(B_{s},\Gamma)=\sigma+k\Delta\sigma^{stat},
\end{equation}

\noindent where $k$ is the statistical factor equal to 1.64 corresponding to 90\% confidence level as given in PDG~\cite{PDG}).
Fig.~\ref{upper_limit_1d} shows the systematic
limits (blue lines) in addition to the statistical uncertainties (green lines).
Systematic errors were estimated by changing the parameters of all cuts applied in the data analysis, and changing the values of assumed potential parameters for the $^3$He-$\eta$ interaction that determines the Fermi momentum distribution for relative motion in the bound state. The highest contribution to the systematic error is connected with the background fit function. The uncertainty due to the fit of quadratic or linear function estimated as $\sigma_{quad}-\sigma_{lin}$ varies from about $2$ to $5~nb$. 

%The uncertainty caused by the fit of quadratic or linear function to the background, estimated as

%}}
%The $\chi^2$ value is then defined as:
%\begin{equationarray}
%\chi^2 = \\
%\sum_i {
%	(\frac{
%		\rho^{measured}_{^3He2\gamma,i}-\rho^{fit}_{^3He2\gamma}(Q_{i})
%	}{
%		\Delta\rho^{measured}_{^3He2\gamma,i}
%	})^2
%} \\
%+	\sum_i {
%	(\frac{
%		\rho^{measured}_{^3He6\gamma,i}-\rho^{fit}_{^3He6\gamma}(Q_{i})
%	}{
%		\Delta\rho^{measured}_{^3He6\gamma,i}
%	})^2
%},
%\label{excitation_curves_fit_chisq}
%\end{equationarray}
%where the $i$ index denotes sum over all $Q_{^3He\eta}$ bins,
%$Q_i$ means $Q_{^3He\eta}$ value corresponding to $i$th bin center,
%$\rho^{measured}_{^3He2\gamma,i}$ and $\rho^{measured}_{^3He6\gamma,i}$ normalized events counts measured for $i$th bin,
%$\Delta\rho^{measured}_{^3He2\gamma,i}$ and $\Delta\rho^{measured}_{^3He6\gamma,i}$ are the statistical uncertainties obtained for $i$th bin,
%and $\rho^{fit}_{^3He2\gamma}(Q_{i})$ and $\rho^{fit}_{^3He6\gamma}(Q_{i})$ are the fitting functions from Eq.~\ref{excitation_curves_fit_function1} and \ref{excitation_curves_fit_function2}.

%The statistical uncertainties for the fit parameters were calculated using the approximation assuming parabolic shape of $\chi^2$ dependence on fit parameters near the minimum position. The upper limit was determined at $90\%$ confidence level (Fig.~\ref{upper_limit_1d}).

%{\it
%
%Magda: please add
%few sentences on using the Hirenzaki code to convert
%the binding energy and width to the potential parameters.
%Interesting that 
%for the potential, before systematics,
%statistical results only OK with QMC prediction of 100 MeV
%potential depth at nuclear matter density \cite{Bass:2005hn},
%and a bit bigger
%than model dependent extractions using He-4 data and 
%the optical model of
%Ikeno et al. \cite{Ikeno:2017xyb}
%most of the model parameter space is excluded allowing 
%values of the real and imaginary parts of the
%potential only between zero and 
%about -60~MeV and -7~MeV
%respectively~\cite{Skurzok_PLB2018}.  
%However, cannot draw any too strong conclusion because of
%the systematics.
%** these few sentences added to wake up interest among
%theoreticians about what might and might not be :-) **
%}

%\textit{\textcolor{blue}{MS: 
In the obtained excitation functions one can see a slight signal from the possible bound state for $\Gamma>20$~MeV and $B_{s}\in[0;15]$~MeV corresponding to the optical potential parameters $-100<V_{0}<-70$ MeV and $|W_{0}|>20$ MeV in the model described in \cite{Skurzok_Hirenzaki2019}. 
The result is also consistent
with the QMC prediction of a
potential depth about -100 MeV at nuclear matter density \cite{Bass:2005hn} and with the models in Refs.~\cite{Barnea:2017epo,Barnea:2017oyk,Xie:2016zhs,kelkar:prl2007}. The allowed $V_{0}$-$W_{0}$ area is however different to those deduced from the $\eta$-$^4$He 
system~\cite{Skurzok_PLB2018} %as well as model dependent extractions 
using the optical model of \mbox{Ikeno et al. \cite{Ikeno:2017xyb}} where most of the model parameter space was excluded allowing values of the real and imaginary parts of the potential only between zero and about -60~MeV and -7~MeV respectively.
%~\cite{Skurzok_PLB2018}.  
%
However, the observed signal is within the range of the
systematic uncertainty. Hence one cannot make definite conclusions whether $\eta$-mesic $^3$He exists with the decay mechanism studied here.

\section{Conclusions}
The analysis of the $pd\rightarrow$ $^{3}\hspace{-0.03cm}\mbox{He}2\gamma$ and $pd\rightarrow$ $^{3}\hspace{-0.03cm}\mbox{He}6\gamma$ reactions has been performed in order to search for the existence of an $\eta$-mesic $^3$He state.
The analysis of the obtained excitation functions for the $pd\rightarrow$ $^{3}\hspace{-0.03cm}\mbox{He}2\gamma$ and $pd\rightarrow$ $^{3}\hspace{-0.03cm}\mbox{He}6\gamma$ reactions shows slight indication of
the signal from the bound state for $\Gamma>20$~MeV and $B_{s}\in[0;15]$~MeV. 
However, the observed indication is within the range of the systematic error which does not allow one to make a definite conclusion on a possible bound state formation.
%in the considered processes.}} 

The upper limit for the cross section of the bound state production varies between $2$ and $15$~nb depending on the bound state parameters.
It is however important to stress that the determined upper limit concerns the production of the $(^3\mbox{He}$-$\eta)_{bound}$ state and its subsequent disintegration via decay of the $\eta$ meson. The branching ratio for the latter in the nuclear medium remains to be estimated theoretically.

This is the first result obtained for the direct decay of bound $\eta$ meson.
The upper limit is much lower than the limit of $70~nb$ for \mbox{$pd\rightarrow$ ($^{3}\hspace{-0.03cm}\mbox{He}$-$\eta)_{bound} \rightarrow$ $^{3}\hspace{-0.03cm}\mbox{He}\pi^{0}$} 
reaction obtained by the \mbox{COSY-11} Collaboration~\cite{light_cosy11_prev}
and is comparable with the upper limits obtained for the 
$dd\rightarrow$ ($^{4}\hspace{-0.03cm}\mbox{He}$-$\eta)_{bound} \rightarrow$ $^{3}\hspace{-0.03cm}\mbox{He}n\pi^{0}$ and $dd\rightarrow$ ($^{4}\hspace{-0.03cm}\mbox{He}$-$\eta)_{bound} \rightarrow$ $^{3}\hspace{-0.03cm}\mbox{He}p\pi^{-}$reactions~\cite{Adlarson:2016dme}.
The much improved constraint will help tuning theoretical modelling of the $\eta$-nucleon and $\eta$-nucleus
interactions.

%$\eta$-nucleon interaction is what one would like to pin down from experiment. The task would have been easier of course if the existence of an $\eta$ mesic bound state as predicted by one of the models would have been found. In the absence of such a finding, our experimental data is still useful as future theoretical calculations of these reactions could narrow the uncertainty in the understanding of the $\eta$-nucleon (and hence $\eta$-nucleus) interaction.}

%\textit{** and of same order the WASA results for Helium-4}

\section*{Acknowledgements}
We acknowledge the support from the Polish National Science Center through grant No. 2016/23/B/ST2/00784, and from the Foundation for Polish Science through the MPD and TEAM POIR.04.04.00-00-4204/17 programmes. Theoretical parts of this work was partly supported by the Faculty of Science, Universidad de los Andes, Colombia, through project number P18.160322.001-17,
and by JSPS KAKENHI Grant Numbers JP16K05355 (S.H.) in Japan.

\section*{References}

\bibliographystyle{unsrt}
\end{document}